\documentclass[twocolumn,english,showpacs]{revtex4}
\usepackage{amsmath,amssymb,dcolumn}
\usepackage[dvips]{graphics}
\def\imo{i}
\begin{document}
\title{Decay of a charged scalar and Dirac fields in the Kerr-Newman-de Sitter background}
\author{R. A. Konoplya}\email{konoplya@fma.if.usp.br}
\affiliation{Instituto de F\'{\i}sica, Universidade de S\~{a}o Paulo \\
C.P. 66318, 05315-970, S\~{a}o Paulo-SP, Brazil}
\author{A. Zhidenko}\email{zhidenko@fma.if.usp.br}
\affiliation{Instituto de F\'{\i}sica, Universidade de S\~{a}o Paulo \\
C.P. 66318, 05315-970, S\~{a}o Paulo-SP, Brazil}
\begin{abstract}
We find the quasinormal modes of the charged scalar and Dirac
fields in the background of the rotating charged black holes,
described by the Kerr-Newman-de Sitter solution. The dependence of
the quasinormal spectrum upon the black hole parameters mass $M$,
angular momentum $a$, charge $Q$, as well as on values of the
$\Lambda$-term and a field charge $q$ is investigated. Special
attention is given to the near extremal limit of the black hole
charge. In particular, we find that for both scalar and Dirac
fields, charged perturbations decay quicker for $q>0$ and slower
for $q<0$ for values of black holes charge $Q$ less than than some
threshold value, which is close to the extremal value of charge
and depend on parameters of the black holes.
\end{abstract}
\maketitle
\section{Introduction}

Quasinormal spectrum of black holes has been extensively
investigated during recent years for a great variety of black hole
backgrounds and fields, because it is an important characteristic
for observation of the gravitational waves \cite{Kokkotas-99},
stability analysis \cite{stability}, AdS/CFT calculations of
temperature Green functions \cite{AdS/CFT_QNMs}. Special attention
has been paid to perturbations of a scalar field
\cite{QNMs-scalar}, as a simplest model, when the influence of the
spin of the field is neglected. When one considers the charged
black hole, the scalar electrodynamic can successfully model the
interaction of the charged field with the electromagnetic
background of the black hole. Therefore the calculation of the
quasinormal modes of charged fields, initiated in
\cite{Konoplya:2002ky} for charged scalar field in the
Reissner-Nordst\"om and dilaton backgrounds, was continued in the
further research
\cite{Zhou:2003ts,Konoplya:2002wt,He:2006jv,Jing:2004zb}.
In particular, in \cite{Konoplya:2002wt} the quasinormal modes of
the massive charged scalar field were found with the WKB accuracy.
In \cite{Konoplya:2002ky} and \cite{Konoplya:2002wt} it was shown
that the quasinormal modes, corresponding to the charged scalar
field, decay quicker than those of the neutral field unless the
black hole charge is larger than some large near extremal value.
This is opposite to the behavior at asymptotically late times,
characterized by the so-called "tail" decay, when the charged
scalar field decays slower, and therefore, dominates at
asymptotically late times \cite{Hod_charged}. Yet, the quasinormal
frequencies calculated for Reissner-Nordstr\"om black holes in
\cite{Konoplya:2002wt,Konoplya:2002ky,Zhou:2003ts}
with the help of the WKB method need better accuracy and cannot be
trusted near the extremal limit, especially for the scalar case,
because the effective potential is frequency dependent in this
case, and WKB equation for the QN frequency must be solved
together with a frequency dependent equation determining the
position of the maximum of the effective potential.

On the other hand, one has much richer physical situation, when
one takes into consideration all the relevant parameters, such as
black hole angular momentum and the cosmological term, i.e. when
one starts from the Kerr-Newman-de Sitter (KNdS) black hole as a
gravitational background. In this paper we achieve both the above
mentioned aims: to find QN modes with very high accuracy, by using
the convergent Frobenius expansion, and to consider the most
general relevant black hole solution of general relativity, KNdS
solution. The latter gives us dependence of the QN spectrum on a
great number of parameters: charge of the black hole $Q$, charge
of the field $q$, normalized angular momentum of the black hole
$a$, the cosmological term $\Lambda$.

The paper is organized as follows: Sec. II gives some basic
formulas on KNdS metric and on radial wave equation for charged
massless scalar and Dirac fields, and also discusses the system of
units we used for showing the QNMs. Sec. III reviews the obtained
numerical results. In the Conclusion we summarize the obtained
results.

\section{Basic formulas}

In the Boyer-Lindquist coordinates the Kerr-Newman-de Sitter
metric has the form \cite{Suzuki:1998vy}
\begin{widetext}
\begin{equation}
ds^2 = -\rho^2
\left(\frac{dr^2}{\Delta_r}+\frac{d\theta^2}{\Delta_\theta}\right)
-\frac{\Delta_\theta \sin^2\theta}{(1+\alpha)^2 \rho^2}
[adt-(r^2+a^2)d\varphi]^2 +\frac{\Delta_r}{(1+\alpha)^2 \rho^2}(dt-a\sin^2\theta d\varphi)^2,
\end{equation}
where
\begin{equation}
\begin{array}{rclcrcl}
\Delta_r&=&(r^2+a^2)\left(1-\alpha r^2/a^2\right)-2Mr+Q^2&\quad&\alpha&=&\Lambda a^2/3,\\
\Delta_\theta&=&1+\alpha\cos^2\theta,&\quad&\rho^2&=&r^2+a^2\cos^2\theta.\\
\end{array}
\end{equation}
The electromagnetic background of the black hole is given by the
four-vector potential
\begin{equation}
A_{\mu}dx^{\mu}
   =-\frac{Qr}{(1+\alpha)^2\rho^2}(dt-a\sin^2\theta d\varphi).
\end{equation}

The charged scalar and Dirac fields in curved space-time are
described by the following equations of motion :
\begin{equation}\label{2}
\Phi^{(0)}_{;\mu\nu} g^{\mu\nu}-i q A_{\mu} g^{\mu\nu}(2 \Phi^{(0)}_{; \nu}-i q A_{\nu}
\Phi^{(0)})-i q A_{\mu; \nu} g^{\mu\nu} \Phi^{(0)} -\frac{1}{6}R\Phi^{(0)}=0, \qquad (R=4\Lambda) \qquad \mbox{\emph{(charged
scalar)}}
\end{equation}
\begin{equation}
\gamma^a e_a^{\mu}(\partial_\mu+\Gamma_\mu+ q A_{\mu})\Phi^{(1/2)}=0,
\qquad \mbox{\emph{(charged
 Dirac)}}
\end{equation}
where $q$ is the charge of particles and $A_{a}$ is the
electromagnetic potential of the background.

Some manipulation with scalar and Dirac equations allow in the
some separable form (see \cite{Suzuki:1998vy} and references
therein). The existence of the Killing vectors $\partial_{t}$,
$\partial_{\phi}$, implies the exponential harmonics of the
following form $\sim e^{-\imo\omega t}$, $\sim e^{\imo m
\phi}$). After the separation of the angular, radial and time variables
$$\Phi^{(s)}(t,r,\theta,\phi) \propto e^{-\imo\omega t}e^{\imo m
\phi}S(\theta)R(r),$$
one can obtain the equation for the angular part
\cite{Suzuki:1998vy}
\begin{eqnarray}
\Bigg\{&& \makebox[-0.7cm]{} \frac{d}{dx} ( 1+\alpha x^2 ) (1-x^2)
\frac{d}{dx}
+ \lambda - s (1-\alpha) + \frac{(1+\alpha)^2}{\alpha} \xi^2
- 2 \alpha x^2 - \frac{(1+\alpha)^2 m^2}{(1+\alpha x^2) (1-x^2)}
- \frac{(1+\alpha) (s^2 + 2smx)}{1-x^2}\nonumber \\
&+& \frac{1+\alpha}{1+\alpha x^2}
\left[ \ 2s (\alpha m - (1+\alpha) \xi) x
- \frac{(1+\alpha)^2}{\alpha} \xi^2 -2 m (1+\alpha) \xi + s^2 \
\right]\Bigg\}S_{s}(x) =0\label{eqn:sx}
\end{eqnarray}
where $s$ is the spin of the field ($s=0, 1/2$ for the scalar and
Dirac field respectively), $x =
\cos
\theta$, $\xi = a
\omega$, and $\lambda$ is the separation constant (for the
non-rotating case $\lambda=\ell(\ell+1)-s(s-1)$, where $\ell\geq
s$ is positive (half)integer multipole number). The angular
equation can be solved numerically with respect to $\lambda$ for
each value of $\omega$ by using the three-term recurrence relation
\cite{Suzuki:1998vy}.

The equation for the radial part is \cite{Suzuki:1998vy}
\begin{equation}
\Bigg\{ \
\Delta_r^{-s}\frac{d}{dr}\Delta_r^{s+1}\frac{d}{dr}
+\frac{1}{\Delta_r}\left(K^2
- is K \frac{d\Delta_r}
{dr} \right)
+4\imo s(1+\alpha)\omega r -\frac{2\alpha}{a^2}(s+1)(2s+1) r^2
+2s(1-\alpha)-2 i s q Q -\lambda \ \Bigg\} R_s(r) = 0,
\label{eqn:Rr}
\end{equation}
where $K = [\omega(r^2+a^2)- am](1+\alpha)- q Q r$.
\end{widetext}

Generally, this equation has five regular singularities: the event
horizon $r_+$, the internal horizon $r_-$, the cosmological
horizon $r^\prime_+$, the spatial infinity and one more
singularity at $r^\prime_-$. One should note, that $r_\pm$ and
$r^\prime_\pm$ are roots of equation $\Delta_r=0$.

In the limit of $\Lambda \rightarrow 0$ one has,
$$r_\pm \rightarrow M\pm\sqrt{M^2-a^2-Q^2}, \qquad \displaystyle{r^\prime_\pm \rightarrow
\pm\frac{a}{\sqrt{\alpha}}}$$
and, therefore, the spacial infinity becomes an irregular
singularity \cite{Suzuki:1998vy}.

By the definition, the QNMs are eigenfrequencies $\omega$ of
(\ref{eqn:Rr}) which satisfy the following boundary conditions (b.c.):
$\psi$ represents purely ingoing waves at the event $r=r_+$
horizon and purely outgoing waves at the cosmological horizon
$r=r^\prime_+$ (or the spacial infinity if $\Lambda=0$).

Now we introduce the new function, which is regular at these two
points if the QNM b.c. are satisfied
\begin{equation}\label{eqn:sing}
y(r)=r^{2s+1}\left(\frac{r-r_+}{r-r_-}\right)^{s+2\imo
K(r_+)/\Delta_r'(r_+)}e^{-\imo B(r)}R(r),
\end{equation}
where $\displaystyle\frac{dB(r)}{dr}=\frac{K}{\Delta_r}$.

The appropriate Frobenius series are
\begin{equation}\label{eqn:Fb}
y(r)=\sum_{n=0}^{\infty}a_n\left(\frac{r-r_+}{r-r_-}\right)^n\left(\frac{1-\rho r_-/r_+}{1-\rho}\right)^n,
\end{equation}
where $\rho=r_+/r^\prime_+$.

Substituting (\ref{eqn:sing}) and (\ref{eqn:Fb}) into
(\ref{eqn:Rr}), one can obtain the $N$-term recurrence relation
for the coefficients $a_i$
\begin{equation}\label{eqn:rrel}
\sum_{j=0}^{min(N-1,i)} c_{j,i}^{(N)}(\omega)\,a_{i-j}=0,\quad
{\rm for}\,\,i>0\,
\end{equation}
where $N$ depends on the black hole parameters. For the particular
Schwarzschild case $N=3$, but for the general case under
consideration $N$ is higher than 3. We decrease the number of
terms in the recurrence relation using the \emph{Gaussian
eliminations}:
\begin{eqnarray}
&&c_{j,i}^{(k)}(\omega) = c_{j,i}^{(k+1)}(\omega),\qquad
{\rm for}\,\,j=0,\,\,\mbox{or}\,\,i<k,\nonumber \\[2mm]
&&c_{j,i}^{(k)}(\omega) =
c_{j,i}^{(k+1)}(\omega)-\frac{c_{k,i}^{(k+1)}(\omega)\,
c_{j-1,i-1}^{(k)}(\omega)}{c_{k-1,i-1}^{(k)}(\omega)}\,.
\nonumber
\end{eqnarray}
After one finds $c_{j,i}^{(3)}$ numerically, he can solve the
equation with \emph{infinite continued fraction} (see
\cite{Zhidenko:2006rs} for more details)
\begin{eqnarray}
&~&
c_{1,n+1}^{(3)}-\frac{c_{2,n}^{(3)}c_{0,n-1}^{(3)}}{c_{1,n-1}^{(3)}-}
\,\frac{c_{2,n-1}^{(3)}c_{0,n-2}^{(3)}}{c_{1,n-2}^{(3)}-}\ldots\,
\frac{c_{2,2}^{(3)}c_{0,1}^{(3)}}{c_{1,1}^{(3)}}=\nonumber\\
\label{CFeq}&~&=\frac{c_{0,n+1}^{(3)}c_{2,n+2}^{(3)}}{c_{1,n+2}^{(3)}-}
\frac{c_{0,n+2}^{(3)}c_{2,n+3}^{(3)}}{c_{1,n+3}^{(3)}-}\ldots\,.\label{invcf}
\end{eqnarray}

Since we can find the separation constant $\lambda$ for each
particular value of $\omega$, (\ref{invcf}) allows to find QNMs
with the desired precision. This technique of the QN spectrum
calculation was proposed by Leaver
\cite{Leaver:1985ax}.
\bigskip
\begin{figure}
\resizebox{0.8\linewidth}{!}{\includegraphics*{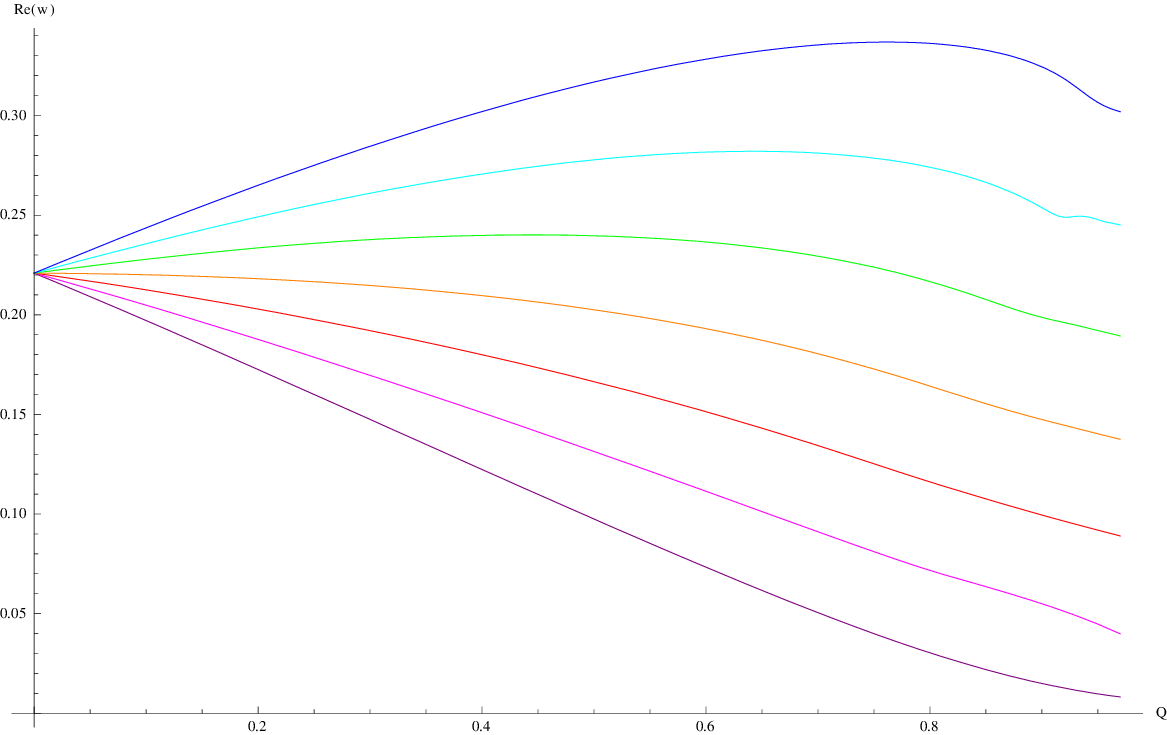}}
\caption{Charged (e) scalar field fundamental frequency ($l=0$) for the
Reissner-Nordstr\"om black hole as a function of its charge ($Q$).
$q=-0.3$ (purple), $q=-0.2$ (magenta), $q=-0.1$ (red), $q=0$
(orange), $q=0.1$ (green), $q=0.2$ (cyan), $q=0.3$ (blue). The
larger $q$ corresponds to the larger real and larger imaginary
(for small black hole charge) part of the QNM.}
\resizebox{0.8\linewidth}{!}{\includegraphics*{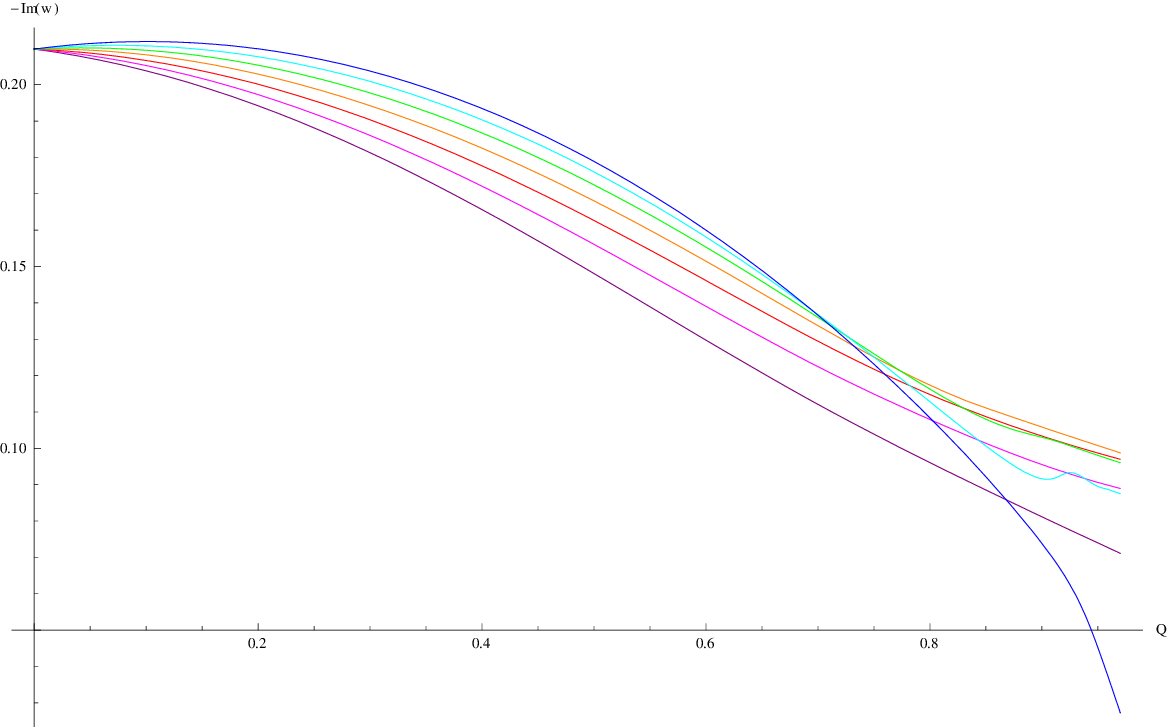}}
\end{figure}

\begin{table*}
\caption{QNMs of the scalar perturbation of the Kerr-Newman black holes ($a=0.4$).}
\begin{tabular}{|r|r|c|c|c|c|c|c|}
\hline
$\ell$&$m$&$Q=0.2$, $q=-0.3$&$Q=0.2$, $q=0$&$Q=0.2$, $q=0.3$&$Q=0.8$, $q=-0.3$&$Q=0.8$, $q=0$&$Q=0.8$, $q=0.3$\\
\hline
$0$&$0$&$0.153622\!-\!0.158460\imo\!$&$0.193444\!-\!0.164777\imo\!$&$0.234254\!-\!0.169836\imo\!$&$0.027107\!-\!0.085597\imo\!$&$0.141475\!-\!0.105014\imo\!$&$0.274019\!-\!0.093648\imo\!$\\
\hline
$1$&$-1$&$0.396131\!-\!0.156125\imo\!$&$0.428770\!-\!0.160382\imo\!$&$0.462269\!-\!0.164345\imo\!$&$0.236903\!-\!0.098374\imo\!$&$0.337666\!-\!0.109255\imo\!$&$0.452137\!-\!0.115708\imo\!$\\
$1$&$0$&$0.479121\!-\!0.151767\imo\!$&$0.514846\!-\!0.154723\imo\!$&$0.551295\!-\!0.157438\imo\!$&$0.285439\!-\!0.093609\imo\!$&$0.399817\!-\!0.098162\imo\!$&$0.528744\!-\!0.097728\imo\!$\\
$1$&$1$&$0.603295\!-\!0.145951\imo\!$&$0.642846\!-\!0.147968\imo\!$&$0.682964\!-\!0.149826\imo\!$&$0.364922\!-\!0.075864\imo\!$&$0.505807\!-\!0.068374\imo\!$&$0.667302\!-\!0.058104\imo\!$\\
\hline
$2$&$-2$&$0.656529\!-\!0.156419\imo\!$&$0.688008\!-\!0.159179\imo\!$&$0.720043\!-\!0.161832\imo\!$&$0.443727\!-\!0.102703\imo\!$&$0.543278\!-\!0.109491\imo\!$&$0.651446\!-\!0.114676\imo\!$\\
$2$&$-1$&$0.726943\!-\!0.154996\imo\!$&$0.760246\!-\!0.157287\imo\!$&$0.794064\!-\!0.159477\imo\!$&$0.488765\!-\!0.100428\imo\!$&$0.595790\!-\!0.105290\imo\!$&$0.711645\!-\!0.108424\imo\!$\\
$2$&$0$&$0.815194\!-\!0.152025\imo\!$&$0.850587\!-\!0.153847\imo\!$&$0.886446\!-\!0.155578\imo\!$&$0.546533\!-\!0.095406\imo\!$&$0.663393\!-\!0.097610\imo\!$&$0.789555\!-\!0.097946\imo\!$\\
$2$&$1$&$0.927889\!-\!0.147965\imo\!$&$0.965608\!-\!0.149382\imo\!$&$1.003731\!-\!0.150726\imo\!$&$0.625000\!-\!0.085165\imo\!$&$0.756010\!-\!0.083476\imo\!$&$0.897414\!-\!0.079871\imo\!$\\
$2$&$2$&$1.072198\!-\!0.144707\imo\!$&$1.112314\!-\!0.145847\imo\!$&$1.152765\!-\!0.146933\imo\!$&$0.742952\!-\!0.064746\imo\!$&$0.897860\!-\!0.058622\imo\!$&$1.064754\!-\!0.052886\imo\!$\\
\hline
\end{tabular}
\end{table*}
\begin{table*}
\caption{QNMs of the scalar perturbation of the Kerr-Newman black holes ($Q=0.2$).}
\begin{tabular}{|r|r|c|c|c|c|c|c|}
\hline
$\ell$&$m$&$a=0.2$, $q=-0.3$&$a=0.2$, $q=0$&$a=0.2$, $q=0.3$&$a=0.8$, $q=-0.3$&$a=0.8$, $q=0$&$a=0.8$, $q=0.3$\\
\hline
$0$&$0$&$0.167766\!-\!0.184220\imo\!$&$0.211875\!-\!0.192171\imo\!$&$0.257011\!-\!0.198680\imo\!$&$0.105432\!-\!0.103623\imo\!$&$0.132795\!-\!0.106936\imo\!$&$0.160904\!-\!0.109273\imo\!$\\
\hline
$1$&$\!-\!1$&$0.470255\!-\!0.176446\imo\!$&$0.507611\!-\!0.180941\imo\!$&$0.545869\!-\!0.185127\imo\!$&$0.266598\!-\!0.109174\imo\!$&$0.289229\!-\!0.112303\imo\!$&$0.312498\!-\!0.115211\imo\!$\\
$1$&$0$&$0.522144\!-\!0.175672\imo\!$&$0.561332\!-\!0.179498\imo\!$&$0.601338\!-\!0.183046\imo\!$&$0.351193\!-\!0.097777\imo\!$&$0.377110\!-\!0.099006\imo\!$&$0.403534\!-\!0.100059\imo\!$\\
$1$&$1$&$0.585319\!-\!0.175043\imo\!$&$0.626523\!-\!0.178296\imo\!$&$0.668457\!-\!0.181312\imo\!$&$0.545520\!-\!0.054487\imo\!$&$0.578091\!-\!0.054264\imo\!$&$0.610920\!-\!0.054055\imo\!$\\
\hline
$2$&$\!-\!2$&$0.787475\!-\!0.176507\imo\!$&$0.823858\!-\!0.179350\imo\!$&$0.860824\!-\!0.182078\imo\!$&$0.438203\!-\!0.109596\imo\!$&$0.459854\!-\!0.111675\imo\!$&$0.481918\!-\!0.113676\imo\!$\\
$2$&$\!-\!1$&$0.834779\!-\!0.176198\imo\!$&$0.872296\!-\!0.178790\imo\!$&$0.910368\!-\!0.181272\imo\!$&$0.505143\!-\!0.106753\imo\!$&$0.528641\!-\!0.108258\imo\!$&$0.552512\!-\!0.109685\imo\!$\\
$2$&$0$&$0.887964\!-\!0.175702\imo\!$&$0.926678\!-\!0.178052\imo\!$&$0.965915\!-\!0.180299\imo\!$&$0.600747\!-\!0.097893\imo\!$&$0.626685\!-\!0.098644\imo\!$&$0.652951\!-\!0.099329\imo\!$\\
$2$&$1$&$0.947908\!-\!0.175148\imo\!$&$0.987872\!-\!0.177273\imo\!$&$1.028326\!-\!0.179304\imo\!$&$0.754274\!-\!0.075327\imo\!$&$0.783768\!-\!0.075164\imo\!$&$0.813525\!-\!0.074974\imo\!$\\
$2$&$2$&$1.015500\!-\!0.174741\imo\!$&$1.056747\!-\!0.176665\imo\!$&$1.098451\!-\!0.178503\imo\!$&$1.039415\!-\!0.048194\imo\!$&$1.072980\!-\!0.048219\imo\!$&$1.106653\!-\!0.048245\imo\!$\\
\hline
\end{tabular}
\end{table*}

Now we shall discuss the units of measurements and ranges of the
black hole parameters. In this paper we shall measure all the
quantities in units of the event horizon. For this we choose the
black hole mass, so that $\Delta_r(1)=0$. Then one has,
$$2M=\left(1-\frac{\Lambda}{3}\right)(1+a^2)+Q^2.$$
We parameterize the cosmological constant $\Lambda$ by the
parameter $\rho<1$
$$\Delta_r\left(1/\rho\right)=0 \quad\Longrightarrow\quad \frac{\Lambda}{3}=
\rho^2\frac{1-\rho(a^2+Q^2)}{1+\rho+\rho^2(a^2+Q^2)}.$$
In these units the condition $\Delta_r'(1)>0$ gives us the range
of values of the black hole charge,
$$Q^2<\frac{1+2\rho}{1+2\rho+3\rho^2+a^2\rho^2}-a^2.$$
The positivity of the righthand side of the above equation bounds
the possible values of the parameter of rotation $a$,
$$a^2<\frac{(1+\rho)\sqrt{1+2\rho+9\rho^2}-(1+2\rho+3\rho^2)}{2\rho^2}\leq1.$$

\section{Results}
\begin{figure}
\resizebox{0.8\linewidth}{!}{\includegraphics*{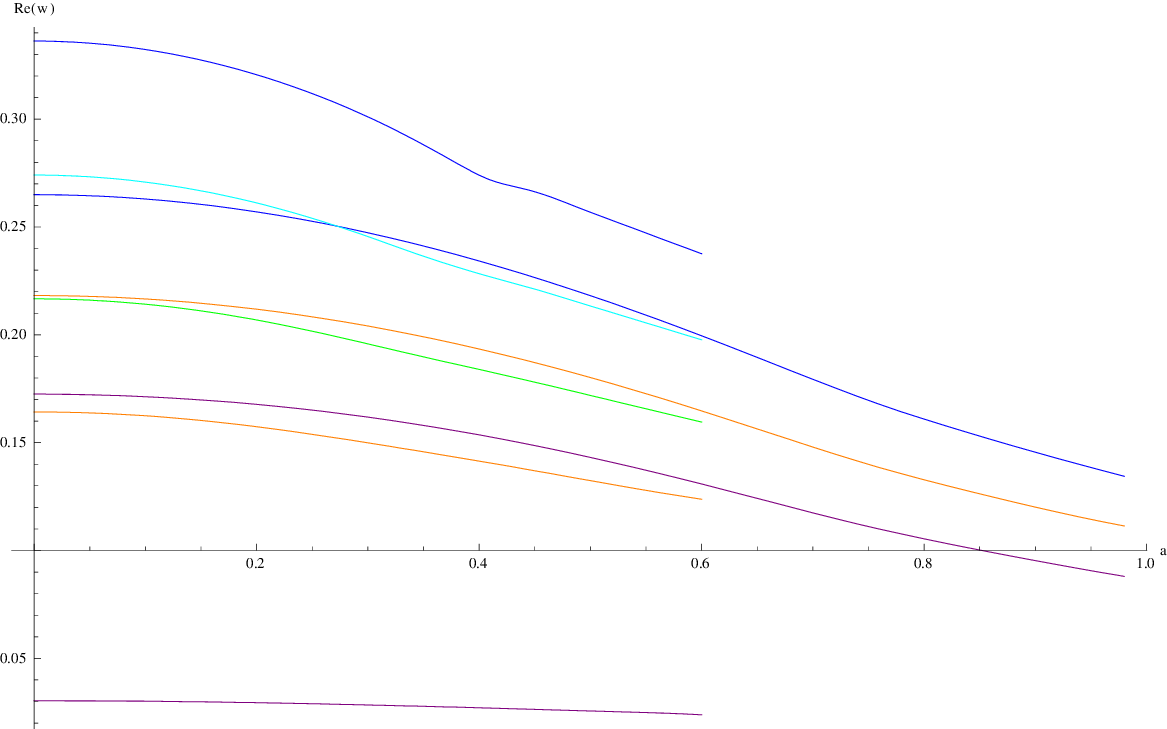}}
\caption{Charged (e) scalar field fundamental frequency ($l=0$) for the Kerr-Newman black hole
of charge $Q=0.2$ ($a<2\sqrt{6}/5$) and $Q=0.8$ ($a<0.6$) as a
function of $a$. $q=-0.3$ $e=0$ (orange), $q=0.1$ (green), $q=0.2$
(cyan), $q=0.3$ (blue). The larger $q$ corresponds to the larger
real and larger imaginary (for small black hole charge) part of
the QNM.}
\resizebox{0.8\linewidth}{!}{\includegraphics*{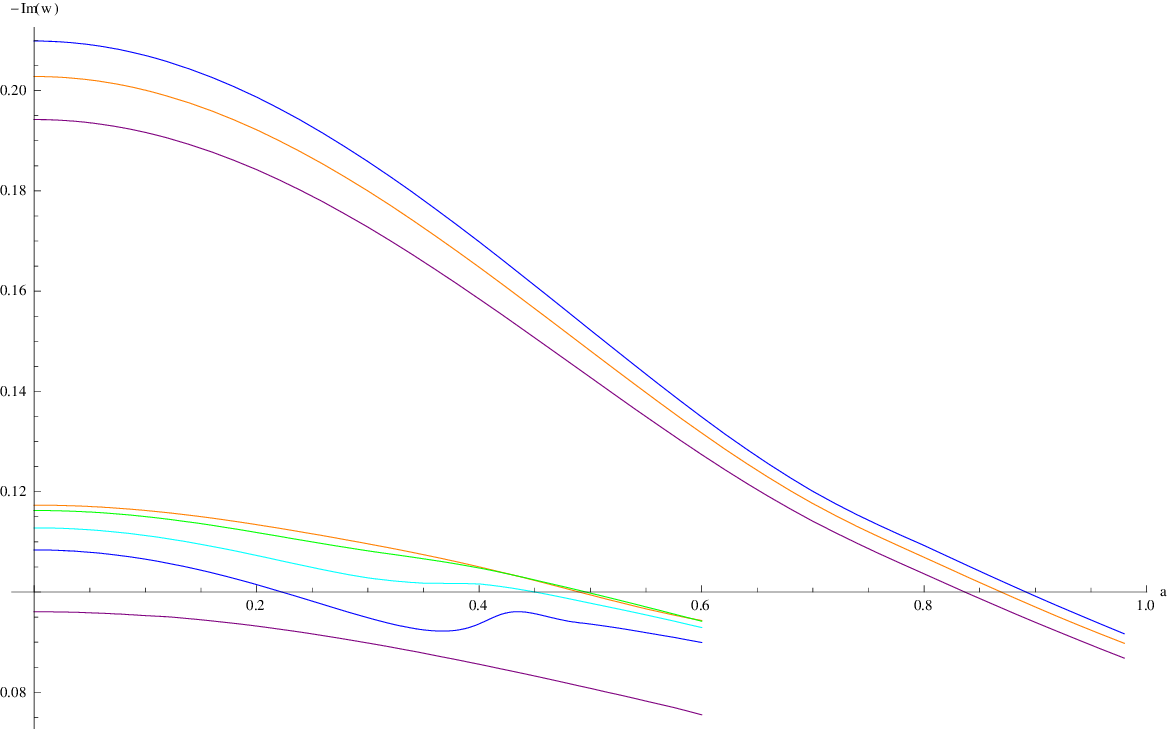}}
\end{figure}
The quasinormal frequencies of the Kerr-Newman-de Sitter black
holes depend on a number of parameters:
\renewcommand{\labelenumi}{\theenumi)}
\begin{enumerate}
\item black hole parameters: mass $M$, charge $Q$, angular momentum $a$,
\item field parameters: field charge $q$ and spin $s$,
\item cosmological constant $\Lambda$,
\item numbers of modes in the spectrum: the multi-pole number $\ell$,
the azimuthal number $m$, and the overtone number $n$.
\end{enumerate}

Therefore, if one wants to represent quasinormal frequencies for
all values of the above parameters, he has to show a great amount
of table data. We decided to be limited here by
\emph{representative} tables or plots, which will show dependence
of the quasinormal modes on each of the above parameters. Thus,
for example on Fig. 1 one can see the dependence of the
$Re(\omega)$ and $Im(\omega)$ on the black holes charge $Q$ for
fundamental mode $\ell=n=0$ of perturbations of the scalar field.
The real part of $\omega$ monotonically grows with the black hole
charge $Q$ and the field charge $q$ (note that $q$ can be both
positive and negative). For positive values of $q$, $Re(\omega)$
attains some maximum value as a function of $Q$, at some large
value of $Q$. Imaginary part has more complicated behavior: it
monotonically decreases as a function of $Q$ until some near
extremal value of the black hole charge, keeping meanwhile
monotonic dependence on $q$. Then the curves with different $q$
intersect, that is, the larger $q$ does not guarantee larger $Im
\omega$. Thus if for not very large $Q$, the charged field
decays quicker than the neutral one, for the near extremal $Q$,
the charged field decays slower than neutral. From Fig. 1 one can
see that this happens at $Q \sim 0.8 Q_{extr}$ in the units of the
event horizon, while in "$M=1$ units" this corresponds to $Q
\sim 0.995 Q_{extr}$. Even though the WKB technique, developed until higher
orders \cite{konoplyaWKB}, reproduces this intersection shown in
Fig 1., it could not be easily trusted in this case because of
$\omega$ dependency of the effective potential
\cite{Konoplya:2002ky}. Therefore, confirmation of the
intersection with the help of the convergent and accurate
Frobenius method leaves apart possible interpretations of charged
quasinormal modes in the context of universality of the critical
collapse \cite{Konoplya:2002wt}. Let us note also that the
non-monotonic behavior for some curves for real and imaginary
parts of $\omega$, near the extremal values of charge, depends on
the value $q Q$ and is not new in fact. When approaching the
extremal limit of values of $Q$ closely enough, one has the
picture of spiraling of the plot of $Re(\omega)$-$Im(\omega)$
\cite{Onozawa}.
\begin{figure}
\resizebox{0.8\linewidth}{!}{\includegraphics*{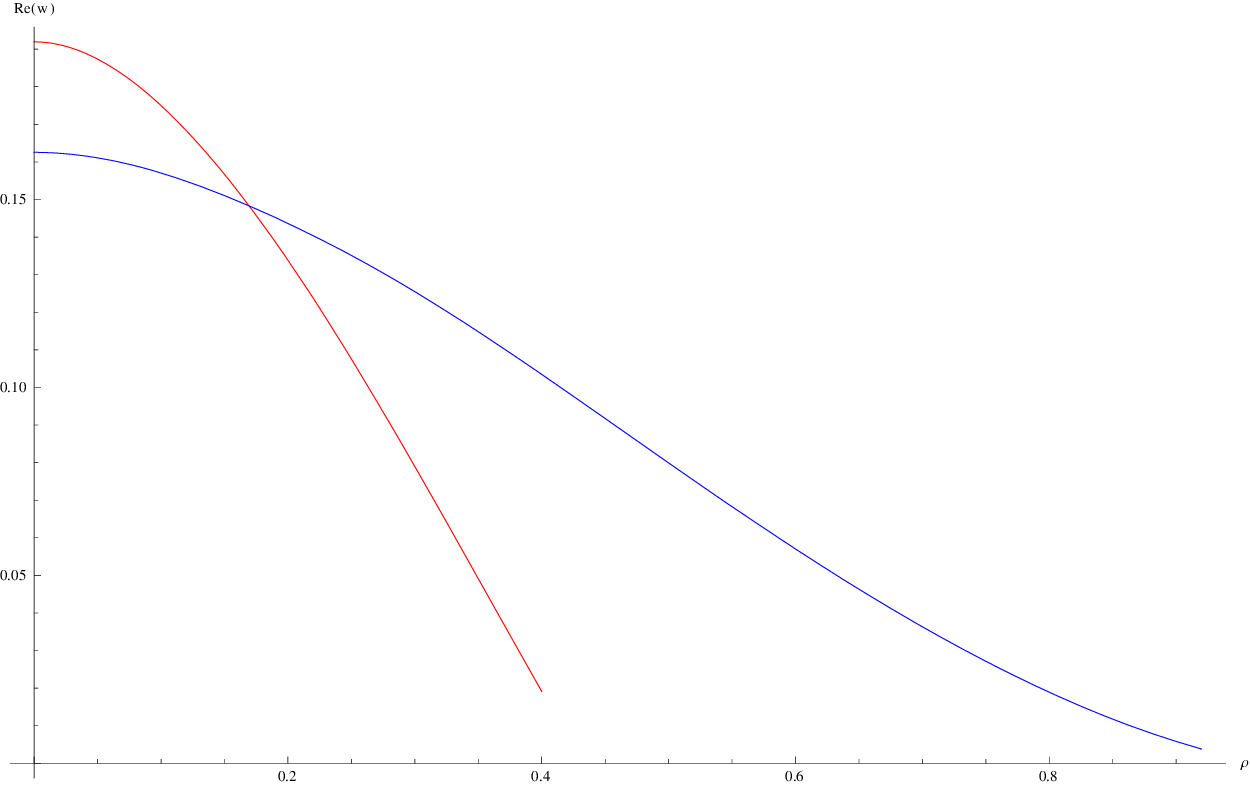}}
\caption{Real and imaginary parts of the fundamental QNMs of Kerr-Newman-de
Sitter BH $s=l=1/2$, $m=-1/2$, $Q=0.2$, $a=0.8$, $q=0$ (red)
($\rho<0.51$) $s=l=m=0$, $Q=0.5$, $a=0.5$, $q=0$ (blue)
($\rho<0.94$ ) as a function of $\rho$.}
\resizebox{0.8\linewidth}{!}{\includegraphics*{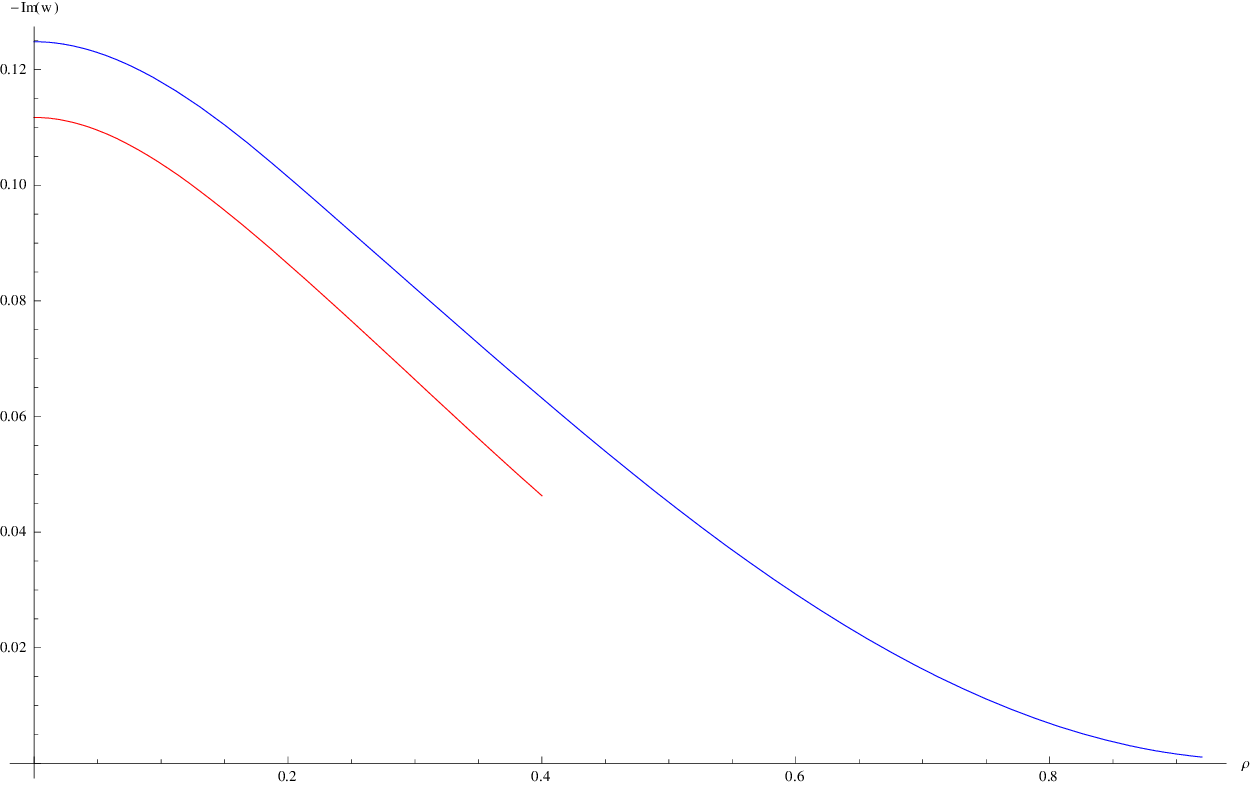}}
\end{figure}

An important point is the checking of all the known particular
limits for our calculations. For $q=Q=0$ we reproduce the
quasinormal modes for the scalar and Dirac fields for Kerr-de
Sitter black holes, while for $\Lambda=0$, $a=0$ we obtain the
limit of pure Reissner-Nordstr\"om black holes and the results of
\cite{Konoplya:2002ky}, \cite{Konoplya:2002wt}. For $q=0$ we find
the quasinormal modes for neutral fields for KNdS black holes.
When approaching the limit of extreme values of the
$\Lambda$-term, one can reproduce the exact quasinormal modes
obtained as a solution of the P\"oschl-Teller equation
\cite{CardosoSdS}, if $q=a=0$.

In Fig. 2 we demonstrate the dependence of $Re(\omega)$ and
$Im(\omega)$ of scalar field modes on the angular momentum $a$ for
a few fixed values of the charges $Q$ and $q$. There one can see
that for not large values of $q Q$, one has the monotonic decay of
both real and imaginary parts of $\omega$ as a function of $a$. In
other words, the quicker rotating black holes have longer lived
modes and smaller real oscillation frequencies. This monotonic
behavior breaks for larger values $q Q$, as it can be seen from
the curve $Q=0.8$, $q=0.3$ on Fig. 2.

In Tables I, III, the quasinormal frequencies for different values
of multi-polar $\ell$ and azimuthal $m$ numbers are given for the
first mode $n=0$, for some values of the charges $Q$ and $q$, and
for a fixed value of the angular momentum. In Tables II, IV, on
the contrary, we fix the values of the black holes charge $Q$, and
consider different values of $a$ and $q$. From the above table one
can see that larger values of the multi-pole number $\ell$, under
the same value $m$, imply larger $Re(\omega)$ and smaller
$Im(\omega)$. Under some fixed $\ell$, large azimuthal numbers
give larger $Re(\omega)$ and smaller damping rates. When the
rotation is large, this decreasing of $Im(\omega)$ for larger $m$
is considerable, so that high $m$ modes can be many times longer
lived than those for non-rotating case $a=0$. This happens for all
values of $Q$ and $q$. Finally, on Fig. 3, one can see the
representative dependence of the quasinormal frequencies on the
values of the $\Lambda$-term. The behavior is qualitatively the
same as for the ordinary Schwarzschild-de Sitter black holes
\cite{Konoplya:2004uk}, that is, the increasing $\Lambda$-term
suppresses considerably the $Re
\omega$ and $Im \omega$, independently on the values of other parameters.
Let us note that when all the parameters $a$, $Q$ and $q$ are
non-vanishing, and the $\Lambda$-term approaches near the extremal
limit, the Frobenius series converges very slowly, so that we
could not reach there extremal limit very closely.

\begin{table*}
\caption{QNMs of the Dirac field perturbation of the Kerr-Newman black holes ($a=0.4$).}
\begin{tabular}{|r|r|c|c|c|c|c|c|}
\hline
$\ell$&$m$&$Q=0.2$, $q=-0.3$&$Q=0.2$, $q=0$&$Q=0.2$, $q=0.3$&$Q=0.8$, $q=-0.3$&$Q=0.8$, $q=0$&$Q=0.8$, $q=0.3$\\
\hline
$\!1\!/\!2\!$&$\!-\!1\!/\!2\!$&$0.246256\!-\!0.154133\imo\!$&$0.280565\!-\!0.160311\imo\!$&$0.316133\!-\!0.165779\imo\!$&$0.118776\!-\!0.090357\imo\!$&$0.221025\!-\!0.107204\imo\!$&$0.344150\!-\!0.113238\imo\!$\\
$\!1\!/\!2\!$&$1\!/\!2\!$&$0.342822\!-\!0.141572\imo\!$&$0.380628\!-\!0.144882\imo\!$&$0.419348\!-\!0.147793\imo\!$&$0.171964\!-\!0.078214\imo\!$&$0.297229\!-\!0.073804\imo\!$&$0.454538\!-\!0.058993\imo\!$\\
\hline
$\!3\!/\!2\!$&$\!-\!3\!/\!2\!$&$0.515027\!-\!0.155523\imo\!$&$0.546793\!-\!0.158973\imo\!$&$0.579262\!-\!0.162249\imo\!$&$0.332623\!-\!0.100337\imo\!$&$0.432175\!-\!0.108952\imo\!$&$0.542690\!-\!0.114981\imo\!$\\
$\!3\!/\!2\!$&$\!-\!1\!/\!2\!$&$0.588694\!-\!0.153027\imo\!$&$0.622736\!-\!0.155715\imo\!$&$0.657409\!-\!0.158246\imo\!$&$0.378403\!-\!0.097191\imo\!$&$0.487664\!-\!0.102514\imo\!$&$0.608266\!-\!0.105015\imo\!$\\
$\!3\!/\!2\!$&$1\!/\!2\!$&$0.687238\!-\!0.148338\imo\!$&$0.723951\!-\!0.150318\imo\!$&$0.761209\!-\!0.152174\imo\!$&$0.442312\!-\!0.088796\imo\!$&$0.565845\!-\!0.089041\imo\!$&$0.701925\!-\!0.086213\imo\!$\\
$\!3\!/\!2\!$&$3\!/\!2\!$&$0.820628\!-\!0.143438\imo\!$&$0.860126\!-\!0.144895\imo\!$&$0.900057\!-\!0.146265\imo\!$&$0.543282\!-\!0.067958\imo\!$&$0.693191\!-\!0.060714\imo\!$&$0.858738\!-\!0.053488\imo\!$\\
\hline
\end{tabular}
\end{table*}
\begin{table*}
\caption{QNMs of the Dirac field perturbation of the Kerr-Newman black holes ($Q=0.2$).}
\begin{tabular}{|r|r|c|c|c|c|c|c|}
\hline
$\ell$&$m$&$a=0.2$, $q=-0.3$&$a=0.2$, $q=0$&$a=0.2$, $q=0.3$&$a=0.9$, $q=-0.3$&$a=0.9$, $q=0$&$a=0.9$, $q=0.3$\\
\hline
$\!1\!/\!2\!$&$\!-\!1\!/\!2\!$&$0.287309\!-\!0.174116\imo\!$&$0.325875\!-\!0.180887\imo\!$&$0.365774\!-\!0.186924\imo\!$&$0.151829\!-\!0.097351\imo\!$&$0.173649\!-\!0.101276\imo\!$&$0.196306\!-\!0.104704\imo\!$\\
$\!1\!/\!2\!$&$1\!/\!2\!$&$0.340964\!-\!0.170124\imo\!$&$0.381353\!-\!0.175322\imo\!$&$0.422875\!-\!0.179958\imo\!$&$0.260575\!-\!0.043041\imo\!$&$0.289055\!-\!0.039671\imo\!$&$0.318597\!-\!0.036871\imo\!$\\
\hline
$\!3\!/\!2\!$&$\!-\!3\!/\!2\!$&$0.615316\!-\!0.175418\imo\!$&$0.651857\!-\!0.178999\imo\!$&$0.689134\!-\!0.182397\imo\!$&$0.311602\!-\!0.098794\imo\!$&$0.331456\!-\!0.101142\imo\!$&$0.351785\!-\!0.103372\imo\!$\\
$\!3\!/\!2\!$&$\!-\!1\!/\!2\!$&$0.663381\!-\!0.174822\imo\!$&$0.701296\!-\!0.178001\imo\!$&$0.739902\!-\!0.181009\imo\!$&$0.378115\!-\!0.093711\imo\!$&$0.400157\!-\!0.095125\imo\!$&$0.422609\!-\!0.096429\imo\!$\\
$\!3\!/\!2\!$&$1\!/\!2\!$&$0.718975\!-\!0.173997\imo\!$&$0.758346\!-\!0.176798\imo\!$&$0.798356\!-\!0.179445\imo\!$&$0.488898\!-\!0.074552\imo\!$&$0.514298\!-\!0.074568\imo\!$&$0.540039\!-\!0.074515\imo\!$\\
$\!3\!/\!2\!$&$3\!/\!2\!$&$0.783304\!-\!0.173201\imo\!$&$0.824167\!-\!0.175660\imo\!$&$0.865612\!-\!0.177982\imo\!$&$0.749656\!-\!0.064353\imo\!$&$0.781326\!-\!0.064171\imo\!$&$0.813053\!-\!0.064004\imo\!$\\
\hline
\end{tabular}
\end{table*}

\section{Conclusion}
In this work, with the help of an accurate convergent Frobenius
method, we performed an extensive calculation of quasinormal modes
of charged scalar and Dirac fields for Kerr-Newman-de Sitter black
holes and have analyzed the dependence of the QN spectrum upon the
great variety of parameters of the black holes $Q$, $M$, $a$, of
$\Lambda$-term, and of the field parameters $q$ and $s$. This
generalizes a number of previous works when only some of the
parameters were considered non-vanishing. The model we considered
might be successful, when considering the interaction of the
charged fields with the electromagnetic background of rotating
black holes.
\begin{acknowledgments}
This work was supported by \emph{Funda\c{c}\~ao de Amparo \`a
Pesquisa do Estado de S\~ao Paulo (FAPESP)}, Brazil.
\end{acknowledgments}



\begin{thebibliography}{99}
\bibitem{Kokkotas-99}  K. D. Kokkotas and B. G. Schmidt,
Living Rev. Relativity \textbf{2}, 2 (1999).
\bibitem{stability}
  R.~A.~Konoplya and A.~Zhidenko,
 Nuclear Physics B777, 182 (2007) [arXiv:hep-th/0703231]
\bibitem{AdS/CFT_QNMs}  G.~T.~Horowitz and V.~E.~Hubeny,
  Phys.\ Rev.\  D {\bf 62}, 024027 (2000)
  [arXiv:hep-th/9909056];
  D.~T.~Son and A.~O.~Starinets,
  arXiv:0704.0240 [hep-th];
  A.~O.~Starinets,
  Phys.\ Rev.\  D {\bf 66}, 124013 (2002)
  [arXiv:hep-th/0207133];
  V.~Cardoso and J.~P.~S.~Lemos,
  Phys.\ Rev.\  D {\bf 63}, 124015 (2001)
  [arXiv:gr-qc/0101052];
  R.~A.~Konoplya,
  Phys.\ Rev.\  D {\bf 68}, 124017 (2003)
  [arXiv:hep-th/0309030];
  V.~Cardoso, R.~Konoplya and J.~P.~S.~Lemos,
  Phys.\ Rev.\  D {\bf 68}, 044024 (2003)
  [arXiv:gr-qc/0305037];
  G.~Michalogiorgakis and S.~S.~Pufu,
  JHEP {\bf 0702}, 023 (2007)
  [arXiv:hep-th/0612065].
  S.~Musiri, S.~Ness and G.~Siopsis,
  Phys.\ Rev.\  D {\bf 73}, 064001 (2006)
  [arXiv:hep-th/0511113];
  I.~Amado, C.~Hoyos, K.~Landsteiner and S.~Montero,
  arXiv:0706.2750 [hep-th].



\bibitem{QNMs-scalar}
  H.~R.~Beyer,
  Commun.\ Math.\ Phys.\  {\bf 221}, 659 (2001)
  [arXiv:astro-ph/0008236];
  L.~E.~Simone and C.~M.~Will,
  Class.\ Quant.\ Grav.\  {\bf 9}, 963 (1992);
  A.~Ohashi and M.~a.~Sakagami,
  Class.\ Quant.\ Grav.\  {\bf 21}, 3973 (2004);
  R.~A.~Konoplya and A.~V.~Zhidenko,
  Phys.\ Lett.\  B {\bf 609}, 377 (2005)
  [arXiv:gr-qc/0411059];
  R.~A.~Konoplya and A.~Zhidenko,
  Phys.\ Rev.\  D {\bf 73}, 124040 (2006)
  [arXiv:gr-qc/0605013];
  A.~Zhidenko,
  Phys.\ Rev.\  D {\bf 74}, 064017 (2006)
  [arXiv:gr-qc/0607133];
  A.~Zhidenko,
  Class.\ Quant.\ Grav.\  {\bf 23}, 3155 (2006)
  [arXiv:gr-qc/0510039];
  E.~Berti and K.~D.~Kokkotas,
  Phys.\ Rev.\  D {\bf 67}, 064020 (2003)
  [arXiv:gr-qc/0301052].
  C.~Ma, Y.~Gui, W.~Wang and F.~Wang,
  arXiv:gr-qc/0611146;
  R.~Konoplya,
  Phys.\ Rev.\  D {\bf 71}, 024038 (2005)
  [arXiv:hep-th/0410057];
  R.~A.~Konoplya and E.~Abdalla,
  Phys.\ Rev.\  D {\bf 71}, 084015 (2005)
  [arXiv:hep-th/0503029];
  R.~A.~Konoplya and R.~D.~B.~Fontana,
  arXiv:0707.1156 [hep-th];
  A.~Lopez-Ortega,
  Gen.\ Rel.\ Grav.\  {\bf 38}, 1565 (2006)
  [arXiv:gr-qc/0605027].




\bibitem{Konoplya:2002ky}
  R.~A.~Konoplya,
  Phys.\ Rev.\  D {\bf 66}, 084007 (2002)
  [arXiv:gr-qc/0207028].

\bibitem{Zhou:2003ts}
  W.~Zhou and J.~Y.~Zhu,
  Int.\ J.\ Mod.\ Phys.\  D {\bf 13}, 1105 (2004)
  [arXiv:gr-qc/0309071].

\bibitem{Konoplya:2002wt}
  R.~A.~Konoplya,
  Phys.\ Lett.\  B {\bf 550}, 117 (2002)
  [arXiv:gr-qc/0210105].

\bibitem{He:2006jv}
  X.~He and J.~Jing,
  Nucl.\ Phys.\  B {\bf 755}, 313 (2006)
  [arXiv:gr-qc/0611003].

\bibitem{Jing:2004zb}
  J.~Jing,
  Phys.\ Rev.\  D {\bf 72}, 027501 (2005)
  [arXiv:gr-qc/0408090].

\bibitem{Hod_charged}
  S.~Hod and T.~Piran,
  Phys.\ Rev.\  D {\bf 58}, 024018 (1998)
  [arXiv:gr-qc/9801001];
  S.~Hod and T.~Piran,
  Phys.\ Rev.\  D {\bf 58}, 024017 (1998)
  [arXiv:gr-qc/9712041];



\bibitem{Suzuki:1998vy}
  H.~Suzuki, E.~Takasugi and H.~Umetsu,
  Prog.\ Theor.\ Phys.\  {\bf 100} (1998) 491
  [arXiv:gr-qc/9805064].
\bibitem{Zhidenko:2006rs}
  A.~Zhidenko,
  Phys.\ Rev.\  D {\bf 74} (2006) 064017
  [arXiv:gr-qc/0607133].
\bibitem{Leaver:1985ax}
  E.~W.~Leaver,
  Proc.\ Roy.\ Soc.\ Lond.\ A {\bf 402}, 285 (1985).
\bibitem{konoplyaWKB} B.F. Schutz and C.M. Will, Astrophys. J. Lett. 291, L33
(1985); S. Iyer and C.M. Will, Phys. Rev. D 35, 3621 (1987);
 R.~A.~Konoplya,
  J.\ Phys.\ Stud.\  {\bf 8}, 93 (2004);
 R.~A.~Konoplya,
  Phys.\ Rev.\  D {\bf 68}, 024018 (2003)
  [arXiv:gr-qc/0303052];


\bibitem{Onozawa} 
  H.~Onozawa, T.~Mishima, T.~Okamura and H.~Ishihara,
  Phys.\ Rev.\  D {\bf 53}, 7033 (1996)
  [arXiv:gr-qc/9603021].

\bibitem{Konoplya:2004uk}
  R.~A.~Konoplya and A.~Zhidenko,
  JHEP {\bf 0406}, 037 (2004)
  [arXiv:hep-th/0402080].
\bibitem{CardosoSdS}
  V.~Cardoso and J.~P.~S.~Lemos,
  Phys.\ Rev.\  D {\bf 67}, 084020 (2003)
  [arXiv:gr-qc/0301078].


\end{thebibliography}
\end{document}